# ALMA and the First Galaxies

Françoise Combes[a]

[a]*Observatoire de Paris, LERMA, CNRS, 61 Av. de l'Observatoire, F-75014 Paris, France*

**Abstract**. ALMA will become fully operational in a few years and open a new window on primordial galaxies. The mm and submm domain is privileged, since the peak of dust emission between 60 and 100 microns is redshifted there for z= 5-10, and the continuum benefits from a negative K-correction. At least 100 times more sources than with present instruments could be discovered, so that more normal galaxies, with lower luminosities than huge starbursts and quasars will be surveyed. The high spatial resolution will suppress the confusion, which plagues today single dish bolometer surveys. Several CO lines detected in broad-band receivers will determine the redshift of objects too obscured to be seen in the optical. With the present instrumentation, only the most massive and gas rich objects have been detected in CO at high z, most of them being ultra-luminous starbursts with an extremely high star formation efficiency. However, selection biases are omni-present in this domain, and ALMA will statistically clarify the evolution of star formation efficiency, being fully complementary to JWST and ELTs.

## ALMA AND THE PRIVILEGED MILLIMETRIC DOMAIN

ALMA with its 54 antennae of 12m and 12 antennae of 7m, will increase the power of millimetric arrays by nearly an order of magnitude. Baselines from 200m to 14km, at wavelengths between 3mm to 0.3mm, will provide spatial resolutions up to 10mas. The large bandwidth of 8GHz/polar will allow to search and determine redshifts. ALMA is not an instrument for big surveys, the field of view is from 1arcmin (at 3mm) to 6 arcsec (at 0.3mm). There is however the possibility of mosaics.

The main advantage for high-z galaxies is that the peak of emission usually around 100 microns is redshifted to the submm and mm. This produces a negative K-correction, i.e. continuum emission from dust is as easy to detect at z=10 than at z=1. In a search of dozens of quasars at redshift between 2 and 6, Beelen (2004) found that about 30% of objects are detected in mm continuum. They correspond to hyper-luminous objects, with L(IR) > $10^{13}$ $L_o$. The derived dust masses of ~$10^8$ $M_o$, mean that dust forms early in the universe.

With ALMA, it will be possible to detect CO lines in a large amount of high-z galaxies, and to search for their redshift, when obscured objects are only detected in their mm continuum. For z>6 galaxies, the high-J CO lines (J>7) will be observed at low frequencies (3mm) with a field of view of 1 arcmin, and a bandwidth of 2x 8GHz ~ 16%, or 50 000km/s. At z=6, the spacing between the various CO lines of the rotational ladder is of 16 GHz, so that with 2 tunings, one obtains a « redshift-machine ».

## Predictions for continuum emission and lines with ALMA

Several studies have computed the expected signals for continuum and CO lines for galaxies at high redshift. Extrapolations of ULIRG ($M(H_2)$ =6 $10^{10} M_o$, $N(H_2)$ = 3.5 $10^{24}$ cm$^{-2}$) shows that it will be easy to detect them in CO until z=3-6 with ALMA, provided that they have reached already solar metallicity in their centre (Combes et al 1999). On the contrary, more normal galaxies, like Lyman-Break (LBG) will be just detected at 3σ and mapped, at z~3, with a spatial resolution of 0.1" (Greve & Sommer-Larsen 2008).

A breakthrough will also occur for molecular absorptions. Up to now, only 5 systems are detected in radio for molecular absorptions (Combes & Wiklind 1997): two are internal absorbers (PKS1413, B3 1504) and three intervening systems, that plays the role of gravitational lenses (B0218, PKS1830, PMN J0134 in OH), since the cross-section of molecular absorbers is very small. There may be 30-100 times more continuum sources detected with ALMA, where absorbers could be detected. They will allow to probe the chemistry of the ISM at high z, and also to check with high precision the variations of fundamental constants (Combes 2009).

Obreschkow et al (2009) have used the millennium simulation, and simple assumptions about the CO emission to derive the CO luminosity function until z=10. They conclude that the detection of regular galaxies, without strong AGN or starbursts, at high z (> 7) in CO emission will be significantly hindered by the weak contrast against the CMB. Although uncertain, it appears that the $H_2$/H I ratio is likely to increase regularly with lookback time, as $H_2/HI \sim (1+z)^{1.6}$, due to the increase of gas density and gas pressure at high z (Obreschkow & Rawlings 2009).

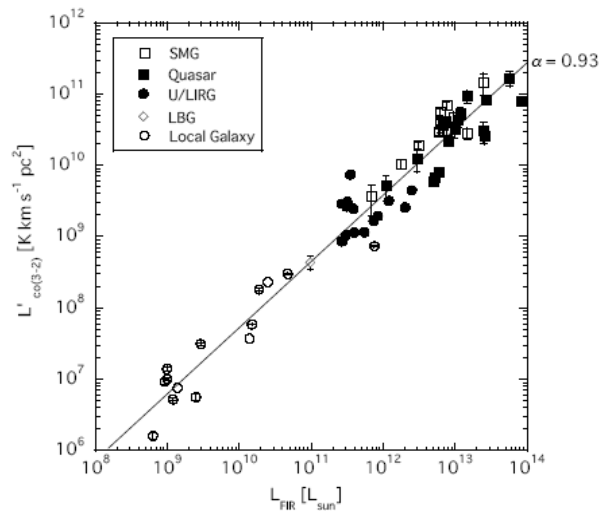

**FIGURE 1.** CO(3-2) luminosity versus FIR luminosity for a large variety of objects, from starbursts to quasars. . From Iono et al (2009).

# REVIEW OF PREVIOUS RESULTS

## CO in high-z galaxies

CO emission is now (2010) detected in about 70 sources at high z. The first historical detection was towards the faint IRAS source F10214+4724 at z=2.3 (Brown & van den Bout 1992, Solomon et al 1992). This was a surprise, but soon it was realized that a high amplification due to an intervening gravitational lens was the reason (Downes et al 1995). Recently Iono et al (2009) have shown how well the CO(3-2) luminosity correlates with the FIR luminosity, even for quasars (cf Figure 1). This suggests that the far-infrared is coming essentially from dust heated by a starburst, frequently associated to an AGN.

Once the continuum is detected in a high z submillimeter galaxy, the CO line detection brings a lot of extra information, in particular the velocity profile can give the dynamical mass, difficult to have from optical lines. An example is SMM J2399-0136 at z=2.808, first detected in CO(3-2) by Frayer et al (1998) and subsequently mapped with the IRAM interferometer by Genzel et al (2003). It benefits from an amplification factor of 2.5, and looks like an inclined disk, with a double-horn spectrum. Its rotation velocity of $\geq 420$ km/s implies a total dynamical mass of $\geq 3 \times 10^{11} \sin^{-2} i$ Mo within an intrinsic radius of 8 kpc, an exceptional mass.

At high z (> 4) detected galaxies are almost all amplified quasars, including the most distant one z=6.4 (Wang et al 2010). This QSO was detected by the SDSS (Fan et al 2003) and is tracing the end of the reionization epoch, as revealed by the Gunn-Peterson trough in its spectrum. The continuum emission yields a dust mass of $M_{dust}$ ~$10^8$Mo (Bertoldi et al 2003), and the massive black hole at its centre has $M_{BH} = 1.5$ $10^9$Mo, according to the BLR spectrum (Willott et al 2003); it is accreting at Eddington rate. No HCN was detected, while the CII line is (Walter et al 2009). With a beam of 0.3", it was possible to constrain the size of the starburst at <1kpc scale, with 1000 Mo/yr/kpc$^2$. This highly concentrated starburst is very similar to local ULIRGs. However, there are some z>4 objects that are more extended, detected in a few lensed objects, 1.4 Gyr after the Big-Bang. BRI 1335-0417 or the Einstein ring, are starbursts of 5kpc extent (Riechers et al 2008a,b). These gas rich mergers have been mapped in CO(2-1) with the EVLA, resolving giant molecular associations, and corresponding brightness temperatures reach 30-40K, close to the true undiluted one. Molecular masses derived are huge, between $10^{10}$ and $10^{11}$ Mo, approaching the dynamical masses in the centre. The black hole masses, estimated from the Eddington limit, are one order of magnitude higher than expected from their bulge mass.

When the interferometer allows to resolve the starburst, in z=4-6 quasars, the warm and highly excited molecular gas, together with the heated dust, reach the maximum surface densities predicted by dust-opacity Eddington limited star formation over kiloparsec scales (Riechers et al 2009a, cf Figure 2). The star formation rates are of ~500-1000 Mo/yr/kpc$^2$. The average surface density $N(H_2)$ ~$10^{24}$cm$^{-2}$ is larger than that of Giant Molecular Clouds (~200Mo/pc$^2$ or $10^{22}$ H$_2$/cm$^2$), as in local ULIRGs.

Going to more moderate star formation rates, the LBG can only be detected when strongly amplified by lenses. A few have been mapped with IRAM interferometer at

z~3 (Coppin et al 2007, Allam et al 2007). The molecular masses are of the order of a few $10^9$ Mo, about 30% of the stellar mass. With the current SFR of ~ 60Mo/yr, the life-time of the starburst without accretion is 40Myr. These objects appear the high-z analogs of LIRGs.

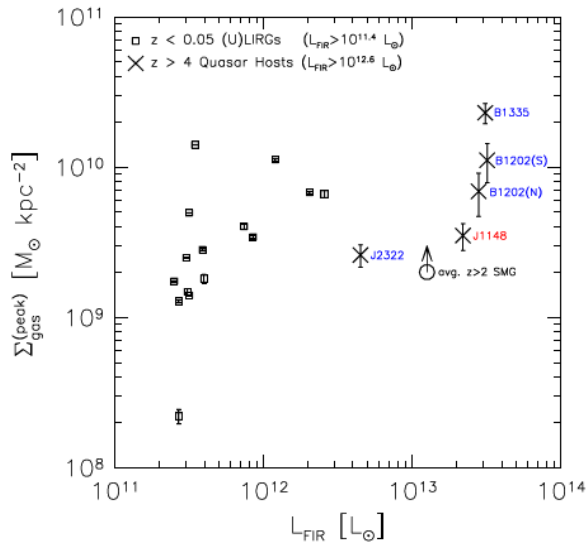

**FIGURE 2.** Molecular gas peak surface density versus FIR luminosity for nearby luminous and ultraluminous infrared galaxies (squares), z > 4 quasar host galaxies (crosses), and for an average of z> 2 SMGs (circle). From Riechers et al (2009a).

## Star formation efficiency in SMG

The star formation efficiency (SFE) can be estimated by the ratio of far-infrared (SF tracer) to CO luminosity (proxy for the amount of gas available). It appears that SFE can be much higher in submillimeter galaxies (SMG) than in ULIRGs. The starburst phase should be as short as 40-200 Myr, with an average SFR ~700 Mo/yr (Greve et al 2005). The reason of this high efficiency is not well known, it could be that the gas fraction is much higher, or dynamical instabilities are more violent (i.e. mergers without bulges).

Recently however another class of ultra-luminous galaxies (L(FIR)~$10^{12}$ Lo), selected by their optical colours, the BzK galaxies, happen to have much more CO emission detected than expected (Daddi et al 2008, cf Figure 3). They are massive galaxies, with large CO extent ~10kpc, with an SFR consumption time scale comparable to the Milky Way, of ~2 Gyr. The gas has low excitation, and the CO lines intensity peak at J=3, like normal galaxies (Dannerbauer et al 2009).

This low excitation suggests that the conversion factor between $H_2$ mass and CO luminosity is standard, i.e. 4.5 times that adopted for ULIRGs. Not all BzK are detected in CO however (Hatsukade et al 2009). There might be a wide range of properties.

Similarly, a recent survey of CO emission in 19 AEGIS galaxies (10 at z~2.3 and 9 at z~1.2) at IRAM interferometer, had surprisingly a high detection rate, larger than 75%, in these « normal » massive Star Forming Galaxies (SFG). The derived gas content is ~34% and 44% in average at z=1.2 and 2.3 respectively (Tacconi et al 2010). This gives some clues to the increase of SFR higher in the past. Either galaxies had a higher gas fraction, or the start formation efficiency was higher. The conclusion is that the first reason is predominant (see Figure 4).

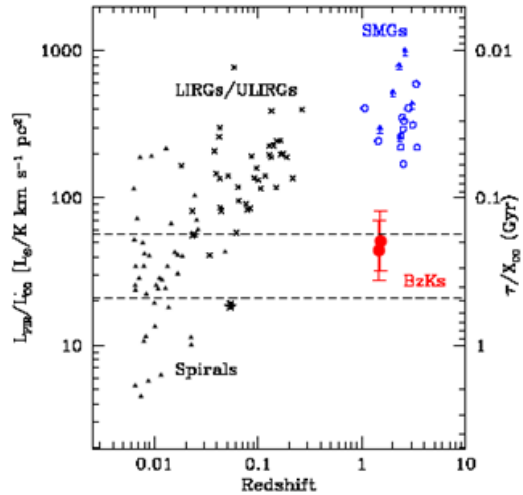

**FIGURE 3.** FIR-to-CO luminosity ratio as a function of redshift. This ratio is a proxy for star formation efficiency. SMGs are located at the top right, and have an exceptional efficiency. The BzK object, by comparison, are almost "normal". The dashed horizontal lines show the semi-interquartile range of local spirals. From Daddi et al (2008).

## THE CO LADDER AND OTHER MOLECULAR TRACERS

The CO lines are the stronger ones, and the best tracer of the molecular component, although HCN appears better correlated to star formation than CO (Gao & Solomon, 2004). Already the 8GHz bandwidth of IRAM receivers (EMIR) at 3mm allows to use the telescope as a redshift-machine. There is always a CO line in the band 80-116GHz, when z>2.2. Weiss et al (2009) have discovered the redshift (2.93) of the submm sources SMMJ14009+0252 with 5 tunings. The CO(3-2) line was detected in the 3mm band and CO(5-4) in the 2mm band. Even when the lines are not consecutive, the redshift is determined unambiguously, since the difference between 2 lines depend on J. To be able to observe z>6 galaxies with ALMA, the CO gas has to be highly excited in the rotational ladder, which might not be the case for moderate starburst however (e.g. Weiss et al 2007). In that case, other tracers of the gas should be searched for, the most obvious being ionised carbon. The CII line unfortunately was not very helpful to study ultra-luminous galaxies, since the CII/FIR ratio drops with luminosity. While for local galaxies it is of the order of $10^{-2.5}$, for ultra-luminous galaxies, it falls to $10^{-3.5}$ and below. However, the CII line could be more useful when more normal galaxies are observed at high z. $H_2$ pure rotational lines from the first galaxies at z=10 will fall in the ALMA high frequency bands, and have been estimated by Spaans & Meijerink (2008), Other elements, such as H3+ or HeH+ are also possible tracers of AGN or starbursts.

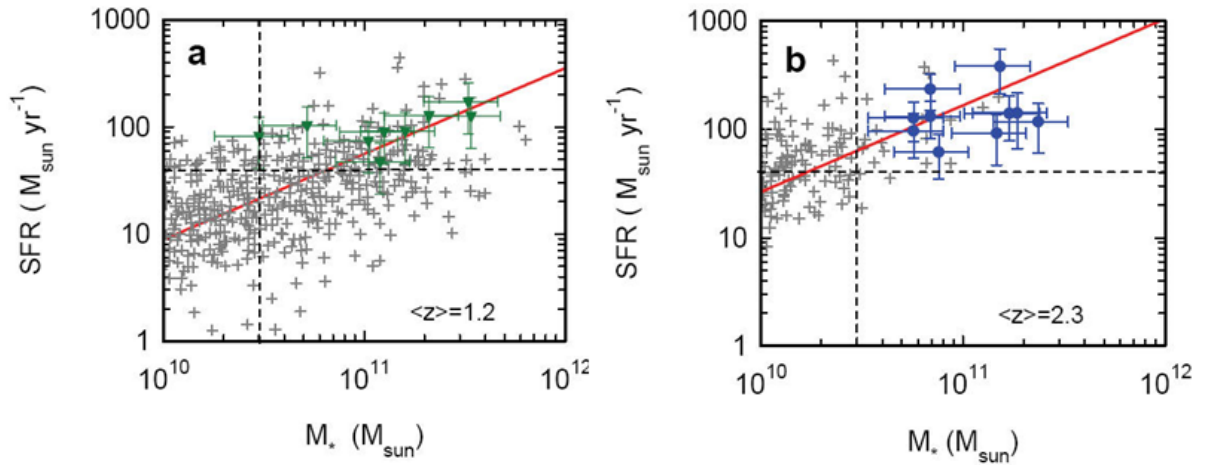

**FIGURE 4.** Star formation rate (SFR) versus stellar mass $M_*$, for the z=1.2 (**Left**), and z=2.3 (**Right**) galaxies. The CO-observed galaxies are marked with the coloured symbols. The best fit (red line) is SFR proportional to $M_*^{0.8} (1+z)^{2.7}$. The slopes of the two red lines are the same (0.8) on the two graphs. The lines derive the one from the other by a vertical translation representing a factor 3, coming from the dependence in (1+z). The latter is mainly due to the gas abundance, and also to a slight variation of the star formation efficiency with time. From Tacconi et al (2010).

## AGN FEEDBACK, STARBURST ASSOCIATION

Many objects detected at high z are active galaxies. The brightest of them, APM08279+5255, is a lensed quasar at z=3.9, with an amplification factor of 50. It has been observed with mm and cm telescopes, and CO lines from J =1 to 11 have been detected. Recent 0.3" resolution CO(1-0) mapping with VLA (Riechers et al 2009b) has shown that CO emission is concentrated in 2 peaks, co-spatial with optical/NIR, and corresponding to the two images of the quasar. The molecules are therefore in a circum-nuclear disk of 550 pc radius, inclined by 25 degrees, with a gas mass of Mgas ~1.3 $10^{11}$ $M_o$. The starburst is quite compact, and does not appear to be quenched by the powerful AGN. The black-hole mass is apparently an order of magnitude larger than expected from the bulge mass, derived from the CO line-width, adopting the above inclination. This surprising black-hole to bulge mass ratio is also observed in a sample of high-z quasars, as shown by Wang et al (2010), see Figure 5. It must be kept in mind that the inclinations of the objects are not well known, and there could be a bias towards detecting less-inclined objects. Also the mass of the black hole could be over-estimated. ALMA will resolve the morphology, and find actual inclinations, and help settling down this issue.

About Lyman-α emitters (LAE) at z>6, there is no detection up to now, neither in continuum nor line. The best upper limit has been obtained towards HCM6A, due to the amplification factor of 4.5, at *z*=6.56 behind the Abell 370 cluster. The flux is lower than 1 mJy at 3σ; the dust mass is lower than 5.3 $10^7$ $M_o$, and SFR < 35 $M_o$/yr (Boone et al 2007). Recently Wagg et al (2009) have obtained a lower limit on the CO emission, with derived $M(H_2)$ < 5 $10^9 M_o$, if the conversion factor is the same as

ULIRGs. However, this conversion factor should be closer to the Milky-Way standard one (factor 4.5), and in the most general case, the amplification factor will not help, so that even ALMA will have difficulties to detect a large number of LAEs.

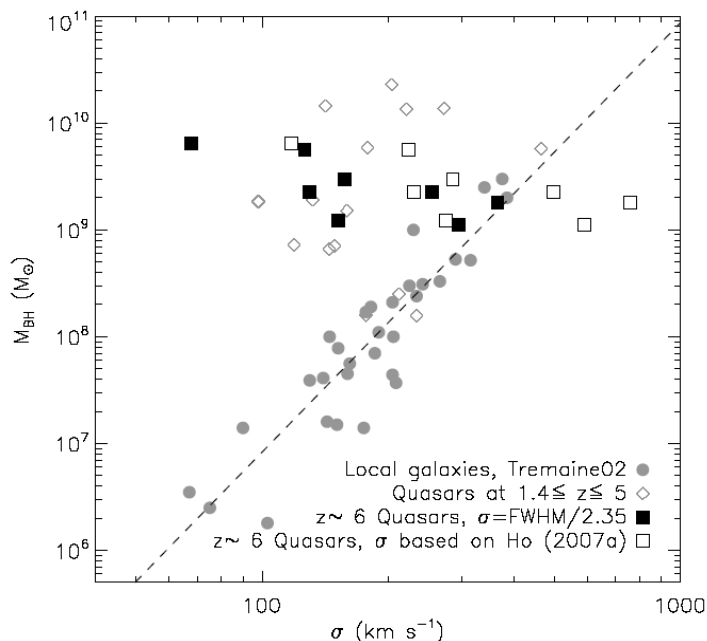

**FIGURE 5.** The high-z quasars appear to have larger black hole masses than what is expected from their bulge velocity dispersion, according to the local relationship (dash line Tremaine et al. 2002). The filled squares and open diamonds are for the z~6 and 1.4 < z <5 quasar samples, respectively, with $\sigma$ derived from the observed CO line width. The open squares show the $\sigma$ values derived with the method by Ho (2007) assuming an average inclination angle of 40° for the z~6 quasars. From Wang et al 2010.

## CONCLUSION AND PERSPECTIVE

With the present generation of mm instruments, it is possible to detect thousands of objects at high redshift in continuum emission, i.e. emission from dust heated by the first star forming galaxies. As for the lines, about 70 systems have been detected in CO at high z. The samples may be dominated by selection effects: only ultra-luminous objects are detected, either with very efficient star formation efficiency (SFE), compact size and highly excited gas, or with more extended gas, normal SFE and SF consumption life-times comparable to the Milky-Way. There is also a bias towards gravitationally lensed objects. In most of detected objects, both AGN and starbursts are contributing to the infra-red emission with varying levels. Quasars are frequently associated to nuclear star formation, which does not appear quenched. It is possible that the black hole masses are larger with respect to the bulge masses, than in local galaxies. With ALMA, much larger samples of primordial galaxies will be discovered on deep field in continuum. The star formation efficiency as a function of redshift will be tackled, while searching for the CO lines, as indicator of H2 content. Dynamical masses can be derived from the CO profiles. When CO high-J lines are not detected, other tracers such as the CII line will trace the interstellar gas, and the low-J CO lines will be searched for with the next generation of cm telescopes (EVLA to SKA).

## ACKNOWLEDGMENTS

I am very grateful to the organizers for their invitation to this exciting meeting, where the occasions of discussion were numerous.